\documentclass[useAMS,usenatbib]{mn2e}

\input{psfig}

\def\xtejdnzh{\mbox{XTE J1908+094}}

\def\cmmoinsdeux{\mbox{ cm}^{-2}}

\def\microns{\mbox{ } \mu \mbox{m}}

\def\kpc{\mbox{ kpc}}

\def\keV{\mbox{ keV}}

\def\asecp{{\rlap.}^{\prime \prime}}

\def\ltsima{\; \buildrel < \over \sim \;}
\def\simlt{\lower.5ex\hbox{\ltsima}}            
\def\gtsima{\; \buildrel > \over \sim \;}
\def\simgt{\lower.5ex\hbox{\gtsima}}            

\usepackage{amssymb,natbib}
\usepackage{graphicx}

\begin{document}

\title[$\xtejdnzh$: Identification of two new NIR candidate counterparts]
{A closer look at the X-ray transient $\xtejdnzh$: 
identification of two new near-infrared candidate counterparts 
\thanks{Based on observations obtained at the Canada-France-Hawaii Telescope (CFHT) which is operated by the National Research Council of Canada, the Institut National des Science de l'Univers of the Centre National de la Recherche Scientifique of France, and the University of Hawaii.}}

\author[S. Chaty et al.]
{S.~Chaty $^1$, R.P.~Mignani $^2$, G.L.~Israel $^3$ \\
%
%
$^1$ AIM - Astrophysique Interactions Multi-\'echelles 
(Unit\'e Mixte de Recherche 7158 
CEA/CNRS/Universit\'e Paris 7 Denis Diderot), \\
CEA Saclay, DSM/DAPNIA/Service d'Astrophysique, B\^at. 709,
L'Orme des Merisiers, FR-91 191 Gif-sur-Yvette Cedex, France
\thanks{chaty@cea.fr} \\
$^2$ {ESO, Karl Schwarzschild Str. 2, D-85748 Garching bei M\"unchen,
Germany} \\
$^3$ {INAF - Osservatorio Astronomico di Roma, Via Frascati 33,
I-00040, Monteporzio Catone, Italy} \\
}
\date{Received date; Accepted date}
    \pubyear{2005} \volume{000} \pagerange{1} \twocolumn

\maketitle \label{firstpage}


\begin{abstract}
We had reported in \cite{chaty:2002c} on the near-infrared (NIR)
identification of a possible counterpart to the black hole candidate
$\xtejdnzh$ obtained with the ESO/NTT.  Here, we present new,
follow-up, CFHT adaptive optics observations of the $\xtejdnzh$ field,
which resolved the previously proposed counterpart in two objects
separated by about $0\asecp8$. 
Assuming that both objects are potential candidate counterparts, we
derive that the binary system is a low-mass system with a companion
star which could be either an intermediate/late type (A-K) main
sequence star at a distance of 3-10 kpc, or a late-type ($>$K) main
sequence star at a distance of 1-3 kpc. 
However, we show that the
brighter of the two objects ($J \sim 20.1$, $H \sim 18.7$, $K' \sim
17.8$) is more likely to be the real counterpart of the X-ray
source. Its position is more compatible with our astrometric solution,
and colours and magnitudes of the other object are not consistent with
the lower limit of 3 kpc derived independently from the peak
bolometric flux of $\xtejdnzh$.  Further multi-wavelength observations
of both candidate counterparts are crucial in order to solve the
pending identification.
\end{abstract}



\section{Introduction} \label{introduction}

The X-ray transient $\xtejdnzh$ was serendipitously discovered on 2002
February 21  during observations  of the field  of the  Soft Gamma-ray
Repeater  SGR 1900+14  performed with  the Proportional  Counter Array
(PCA) aboard RXTE (galactic  coordinates  $l,b$  =  $43.26\deg,
+0.43\deg$; \citealt{woods:2002}).   Within one month after its
discovery, the  source flux increased by  a factor $\sim  3$ and broad
peak QPOs  at a frequency  of 1 Hz  were also detected.   The RXTE/PCA
2-$30  \keV$ spectrum was  found consistent  with a  power-law (photon
index=1.55) with an hydrogen column  density $N_H = 2.3 \times 10^{22}
\cmmoinsdeux$ \citep{woods:2002}.
Soon  after,  $\xtejdnzh$ was  observed  at  higher energies  ($15-300
\keV$) with  the PDS aboard  {\it BeppoSax} \citep{feroci:2002},  at a
flux level  inconsistent with the extrapolation of  the RXTE/PCA power
law, implying a cut-off in  the X-ray spectrum at energies around $100
\keV$.  Target  of Opportunity  (ToO) observations performed  with the
MECS  on  {\it  BeppoSax}  \citep{in'tzand:2002a}  provided  a  source
positioning better than $20\arcsec$, while the measurement of the source
flux  in  outburst  set  a  lower  limit on  the  distance  of  3  kpc
\citep{in'tzand:2002b}.  The transient radio counterpart to $\xtejdnzh$
was identified  with the  Very Large Array  (VLA), with  a position
uncertainty  of  $0.1 \arcsec$  in  each coordinate  \citep{rupen:2002},
consistent     with     the     improved     Chandra/ACIS     position
\citep{in'tzand:2002b}.   First follow-up optical  observations failed
to  detect  any  possible   counterpart  brighter  than  $R  \sim  23$
\citep{wagner:2002}   and   $I   \sim  22$   \citep{garnavich:2002}.
Thereafter,  \cite{chaty:2002a}  reported the  discovery  of a  likely
near-infrared (NIR)  counterpart, apparently variable  along the X-ray
lightcurve \citep{chaty:2002c}. \\ Here, we  report on the results
of  new,  follow-up,  adaptive   optics  NIR  observations  of  the
$\xtejdnzh$ candidate counterpart.
The  observations are described  in Section  \ref{observations}, while
the results are discussed in Section \ref{results}.

\section{Observations and Data Reduction} \label{observations}



The first NIR  observations of the $\xtejdnzh$ field  were obtained in
 April  and  June 2002  with  the  SOFI camera  at  the  ESO 3.5m  New
 Technology Telescope  (NTT) and are  described in \cite{chaty:2002c}.
 New NIR observations were performed on 2002 August 18 in visitor mode
 with the  3.6m Canada-France-Hawaii Telescope (CHFT) on  top of Mauna
 Kea (Hawaii).   The observations were carried out  using the Adaptive
 Optics  Bonnette (PUEO), mounted  at the  f/8 CFHT  Cassegrain focus,
 equipped  with the KIR  instrument. This  is a  near-infrared imaging
 camera based  on the 1024$\times$1024 Rockwell  Science Center HAWAII
 (HgCdTe Astronomical  Wide Area Infrared Imaging)  focal plane array,
 sensitive to radiation from 0.7  to 2.5 $\microns$.  The camera plate
 scale  is $0\asecp035$/pixel,  yielding  a total  field-of-view of  
$36\arcsec \times 36\arcsec$.

Images were taken through the $J$ ($\lambda = 1.250 \microns$, $FWHM =
0.160  \microns$), $H$  ($\lambda  = 1.635  \microns$,  $FWHM =  0.290
\microns$) and K' ($\lambda=2.120 \mu$m,  $FWHM = 0.340 \mu$m) filters.
The source was observed through a sequence of 5 exposures of 90s each
in the $J$, $H$ and $K'$ filters, respectively, with a random offset among
images in order to perform background subtraction of the variable IR
sky.  Conditions were
photometric. The airmass was between 1.02 and 1.04, the typical seeing
conditions were around $0.4\arcsec-0.6\arcsec$ and the adaptive optics
resulted in a source Point Spread Function (PSF) of about
$0\asecp1-0\asecp14$ in the images (diffraction limited).
The images have been corrected for instrumental
effects,  by removal  of  the dark  current  and flat-fielding, 
and then sky subtracted, using
standard procedures  available in the IRAF package  {\it ccdred}.  For
each filter, we finally computed a  final image which is the median of
the  sequence.   Photometric   calibration  was  performed  using  the
standard star field FS35 (G21-15; \citealt{casali:1992}).


\section{Results} \label{results}

    \subsection{Astrometry}

To accurately register the radio coordinates of the source
\citep{rupen:2002} on the CFHT images, we have computed the image
astrometry using as a reference the positions of stars selected from
the Two Micron All Sky Survey (2MASS) Point Source Catalogue, which
has an intrinsic absolute astrometric accuracy of $\sim 0\asecp2$
per
coordinate\footnote{http://spider.ipac.caltech.edu/staff/hlm/2mass/overv/overv.html}.
Due to the smaller field of view of the CFHT camera ($36\arcsec \times
36\arcsec$), only 11 2MASS objects have been identified in the
averaged $J$-band image and used as suitable astrometric calibrators.
The pixel coordinates of the reference stars have been computed by a
two-dimensional gaussian fitting procedure, and transformation from
pixel to sky coordinates was then computed using the programme
ASTROM\footnote{http://star-www.rl.ac.uk/Software/software.htm},
yielding an rms of $\sim 0\asecp08$ in both Right Ascension and
Declination, which we assume representative of the accuracy of our
astrometric solution.  After taking into account the rms of the
astrometric fit ($\sim 0
\asecp08$), the  absolute  accuracy of the 
2MASS  coordinates ($\sim  0\asecp2$) and  the error  of  the source
radio  coordinates  ($0\asecp1$, \citealt{rupen:2002}),  we  estimated
$\sim  0\asecp24$  (per  coordinate)   to  be  the  final  uncertainty
on the computed source position.

\begin{figure*}
\setlength{\unitlength}{1.0cm}
\begin{picture}(8.5,8.5)(4.5,-9)
\includegraphics{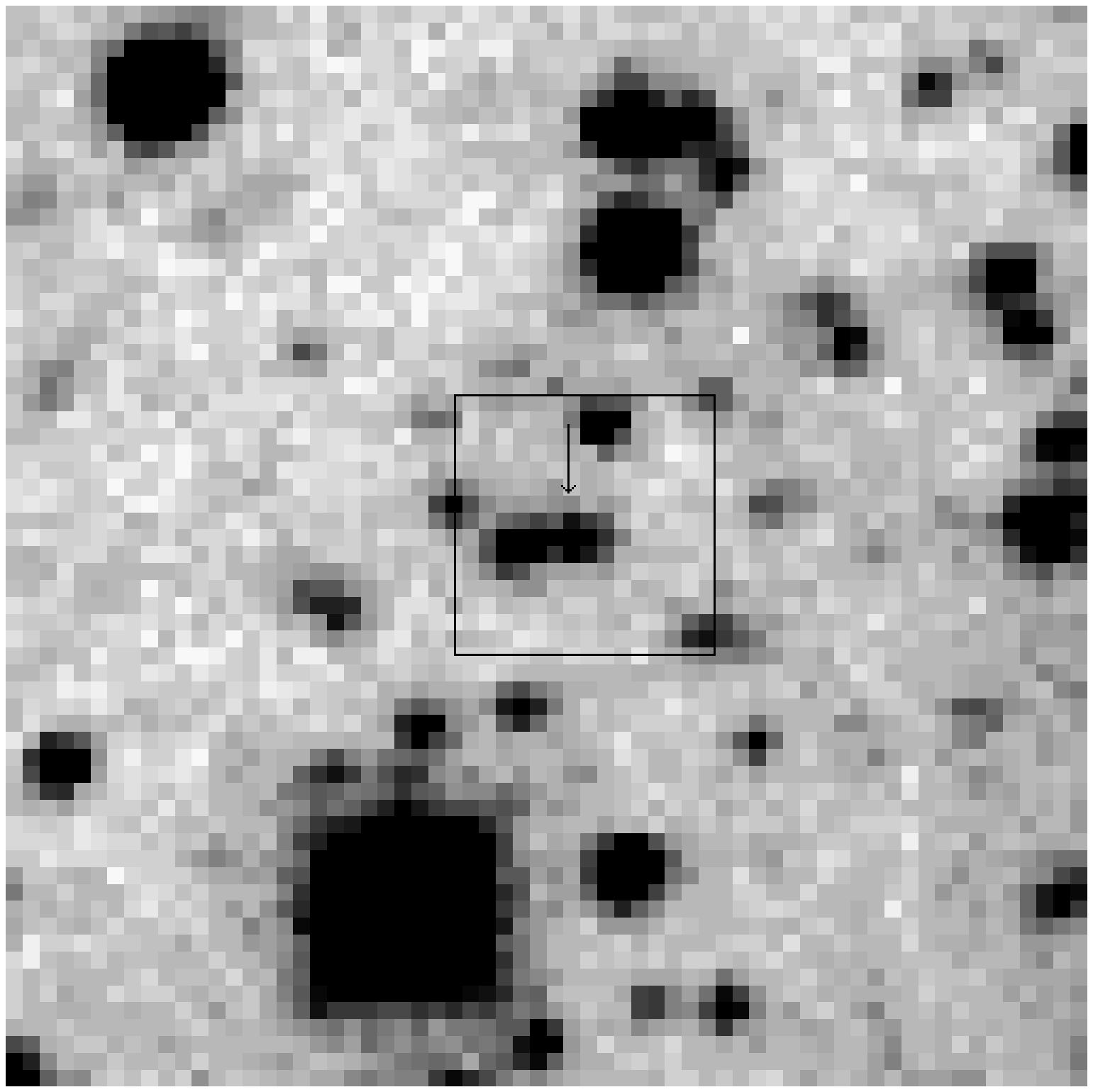}
\end{picture}
\begin{picture}(8.5,8.5)(0.,4.)
\includegraphics{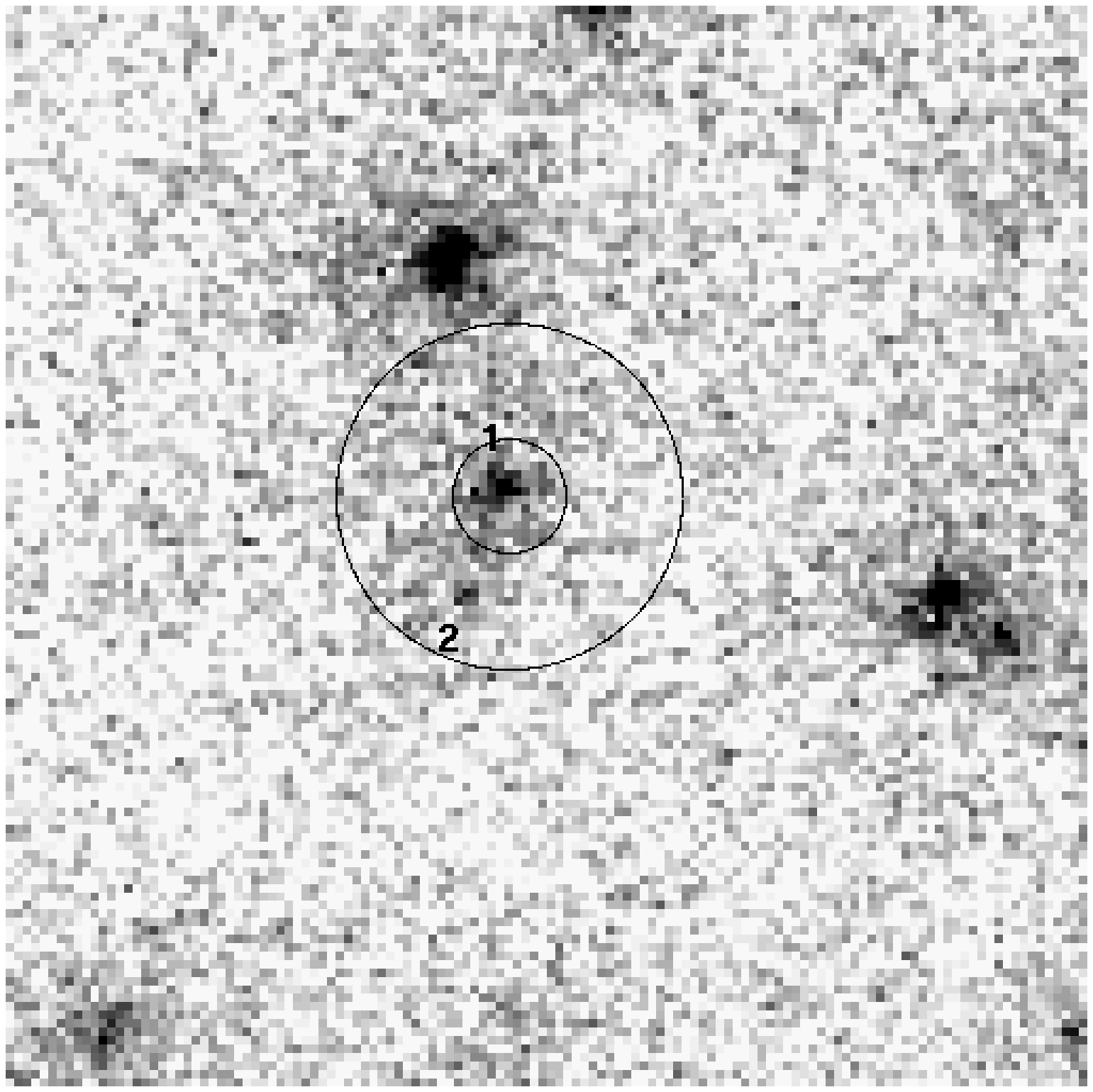}
\end{picture}
\caption[]{\label{xtej1908_fc}    Left panel: $15\arcsec \times 15\arcsec$ 
cutout of the NTT/SOFI $J$-band image of the $\xtejdnzh$ field
\citep{chaty:2002c}, with the previously suggested candidate
counterpart marked by an arrow.  North to the top, East to the left.
Right panel: $4\asecp5 \times 4\asecp5$ cutout (corresponding to
the box in the left panel) of the combined CFHT/KIR $J$ band image of
the same field. In the CFHT image, the two circles (of radius
$0\asecp24$ and $0\asecp71$, respectively) correspond to the $1
\sigma$ and $3 \sigma$ uncertainties on the source position derived from our
astrometric solution and the target radio position ($\alpha(J2000) =
19^{\rm h}08^{\rm m}53\fs77 \pm 0\fs007$; $\delta(J2000)=+09^\circ 23'
04\asecp 9 \pm 0\asecp 1$) from \citet{rupen:2002}.  The two
possible candidate counterparts detected within the $3 \sigma$ error
circle are labelled (1 and 2).  }
\end{figure*}

Figure \ref{xtej1908_fc} shows $15\arcsec \times 15\arcsec$ and
$4\asecp5 \times 4\asecp5$ cutouts of the NTT and CFHT $J$-band images
centered around the computed source position (left and right panel,
respectively). In the CFHT image we overplot the 1 and 3 $\sigma$
error circles derived from our astrometric solution.  One and two objects,
labelled 1 and 2 in figure \ref{xtej1908_fc} (right panel), are found
within the $1 \sigma$ and $3 \sigma$ error circles, respectively.  As
the angular separation between objects 1 and 2 is about $0\asecp8$, we
note that they could have been hardly resolved in the SOFI images of
\cite{chaty:2002c}, in  view  of  the  non-optimal   seeing conditions
($0\asecp8-1\asecp2$) during the observations and of the much larger
pixel size of the detector ($0\asecp292$).  Thus, we conclude that 
the previously proposed candidate counterpart was actually the
blend of these two objects.

Although both objects are considered as potential counterparts in the following analysis, we note that object 1 (the brighter of the two) is certainly favoured as its position is more consistent with our astrometric solution.


\subsection{Photometry}

We  have   computed  the  $JHK'$  photometry  of   the  two  candidate
counterparts (objects 1 and 2)  identified in the CFHT images. In order
to obtain more  precise measures, we have used  the {\it IRAF/DAOPHOT}
package  for  the  photometry  in crowded  fields.   Specifically,  we
computed a typical  PSF for every image using  the tasks {\it daofind,
phot,  psf,  nstar} and  {\it  substar},  iterating  the procedure  to
minimize the  residuals after  PSF subtraction.  Then,  we ran  a star
subtraction    algorithm    through    the    task    {\it    allstar}
\citep{massey:1992} to remove all objects detected in the neighbourhood
of our candidate counterparts before computing their magnitudes.  The
final CFHT $JHK'$ magnitudes of objects 1 and 2 are reported in rows 4
and 5 of Table \ref{table_log}.

Using  the same  procedure  described above,  we  have reanalyzed  the
original NTT  images of \cite{chaty:2002c} in an  attempt to resolve,
or at  least constrain, the flux  of objects 1  and 2. Unfortunately,
the  highly uncertain  residuals  after the  PSF  subtraction make  it
impossible to  derive reliable magnitude estimates for both objects.
Thus,  we have  taken the  magnitudes reported  in \cite{chaty:2002c},
which correspond to  the blend of objects 1 and 2,  as upper limits on
the actual magnitudes of these two objects. These values are listed in
the first three rows of Table \ref{table_log}.

As no direct evaluation of the variability of objects 1 and 2 is
possible from the current photometry, we have tried to obtain an
indirect piece of evidence by simply comparing the combined flux of
the two objects measured in the CFHT images (Table
\ref{table_log}, last row) with the flux of the blend
measured in the NTT ones (Table \ref{table_log}, first three rows).
As it is seen from Table \ref{table_log}, in the H and K bands the
combined flux of objects 1 and 2 is significantly fainter than the
flux of the blend at the different epochs (it is also true in the
$J$-band, however we point out that the large error attached to the
2002-06-17 measurement hide any variation). In other words, the IR
flux of one of the two objects (or both) varied in the time span
covered by our observations.

\begin{figure*}
\centerline{\psfig{file=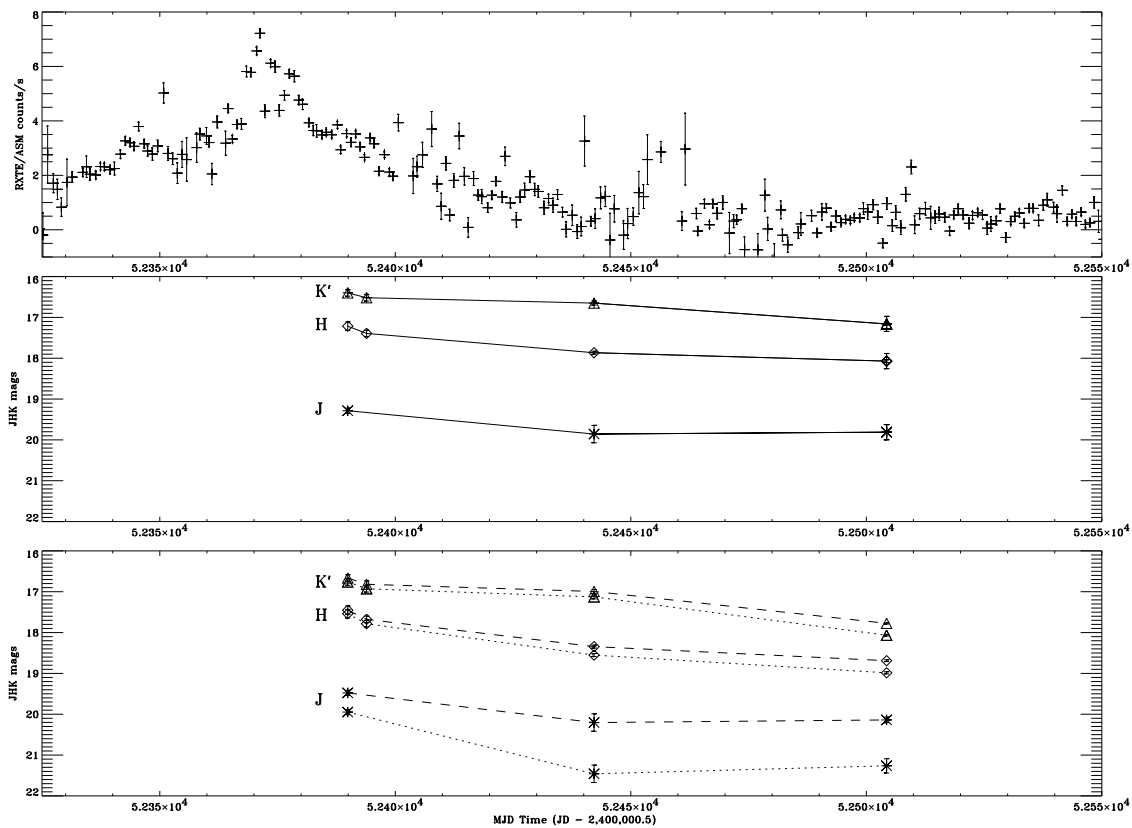,angle=90.,width=15.cm}}
\caption[]{\label{xtej1908_lc}    Multi-wavelength    lightcurve    of
$\xtejdnzh$. Upper panel: {\it RXTE/ASM} $2-10 \keV$ lightcurve of
$\xtejdnzh$. Flux values are in units of counts/s averaged over one
day.
Middle panel: observed IR lightcurves from the $J$ ($\ast$), $H$
($\diamond$) and $K'$ ($\triangle$) band magnitudes listed in
Table \ref{table_log}. Per each passband, the points at the first
three epochs are the magnitudes of the blend of objects 1 and 2
measured with the NTT/SOFI, while the points at the last epoch are the
total magnitudes of objects 1 and 2 measured with the CFHT/KIR.
Lower panel: Same as above, but the couple of points at the first
three epochs are the estimated NTT/SOFI magnitudes for objects 1 and 2,
corresponding to the difference between the total flux of the blend
and the flux of object 2 and 1 (assumed constant), respectively. The
couple of points at the last epoch are the measured CFHT/KIR
magnitudes of objects 1 and 2. The dashed and dotted lines connect
points associated to objects 1 and 2, respectively.
The timescale is the same in the three panels, and the flux scale is
the same in the middle and lower panels, for an easier comparison.
}
\end{figure*}

To search for a possible correlation between the observed IR and X-ray variability, we have compared our photometry reported   in  Table
\ref{table_log} with the RXTE/ASM lightcurve of $\xtejdnzh$ (Figure \ref{xtej1908_lc}, top panel). First, we considered as before the total $JHK$ magnitudes of objects 1 and 2 (Table 1, last row) and the magnitudes of the blend (Table 1, first three rows).  These are plotted in Figure 2 (middle panel). Second, we considered the actual $JHK$ magnitudes of objects 1 and 2 
(Table 1, rows 4 and 5) with the expected magnitudes derived from the
NTT/SOFI upper limits. As a crude approximation, we have straightly
computed (per each passband) the difference between the NTT/SOFI flux
of the blend and the CFHT/KIR flux of objects 1 and 2 in turn,
assuming the other constant. The resulting lightcurves are shown in
Figure \ref{xtej1908_lc} (lower panel). From the IR lightcurves it is
clear that either object 1 or 2 (or both) have undergone a significant
decrease of the NIR flux consistent with the X-ray flux decay which
occurred in the time span covered by our observations.  This
variability means that, almost certainly, one of the two objects is
the real counterpart of the X-ray source $\xtejdnzh$.



\begin{table*}
\begin{center}
\begin{tabular}{|c|c|c|c|c|c|c|}
\hline
 Date      &   MJD   &   Instrument & Object & $J$   & $H$     &  $K'$   
\\
\hline
2002.04.25 & 52389.9 & NTT/SOFI & Objects 1,2 & $>19.29 \pm 0.13$ & $>17.21 \pm 0.21$ & 
$>16.40 \pm 0.16$ 
\\
2002.04.29 & 52393.9 & NTT/SOFI & Objects 1,2 &     -              & $>17.39 \pm 0.16$ & 
$>16.52 \pm 0.16$ 
\\ 
2002.06.17 & 52442.2 & NTT/SOFI & Objects 1,2 & $>19.86 \pm 0.43$ & $>17.86 \pm 0.07$ & 
$>16.65 \pm 0.10$ 
\\
\hline
2002.08.18 & 52504.3 & CFHT/KIR & Object 1 & $20.14 \pm 0.14$  & $18.69 \pm 0.04$ 
& $17.77 \pm 0.03$
\\
2002.08.18 & 52504.3 & CFHT/KIR & Object 2 & $21.26 \pm 0.35$  & $18.98 \pm 0.05$ & 
$18.06 \pm 0.04$ \\
\hline 
2002.08.18 & 52504.3 & CFHT/KIR & Objects 1+2 & $19.81 \pm 0.37$ & $18.07 \pm 0.07$ & $17.16 \pm 0.05$ \\
\hline
\end{tabular}
\end{center}
\caption[]{\label{table_log}  Photometry of the $\xtejdnzh$ candidate 
counterparts. The magnitudes in rows 1-3 were obtained with the SOFI
instrument at the NTT and refer to the original candidate counterpart
reported by \cite{chaty:2002c}, which is actually a blend of two
objects clearly resolved by the CFHT (objects 1 and 2, see Figure
\ref{xtej1908_fc}). Thus, the SOFI magnitudes have to be taken as
upper limits on the magnitudes of both objects 1 and 2.  The magnitudes
in row 4-5 have been obtained with the KIR camera at the CFHT and
refer to objects 1 and 2, respectively.  Finally, the magnitudes in the
last row represent the combined magnitudes of objects 1 and 2 (i.e. an
artificial blending of CFHT objects) in order to compare it to the
SOFI magnitudes.}
\end{table*}


\subsection{The companion star} \label{star}

We have used our $JHK'$ band photometry to derive constraints on the
nature of the two $\xtejdnzh$'s candidate counterparts.  For both
objects 1 and 2, we have taken the magnitudes measured on August 18th
2002, i.e.  likely at the minimum, as inferred from the X-ray
lightcurve (Figure \ref{xtej1908_lc}, upper panel), assuming that all
the flux is produced by the companion star.  However, we have to
caution that this assumption might not be entirely valid, since Figure
\ref{xtej1908_lc} shows that the X-ray source still exhibited some
level of activity at this epoch.  Therefore, it is likely that the
infrared emission was not produced purely from the companion star but
was contaminated by the contribution of the accretion disc. For this
reason, we have prudently assumed the derived absolute magnitudes as
upper limits on the luminosity of the companion star.

\begin{figure}
\centerline{\psfig{file=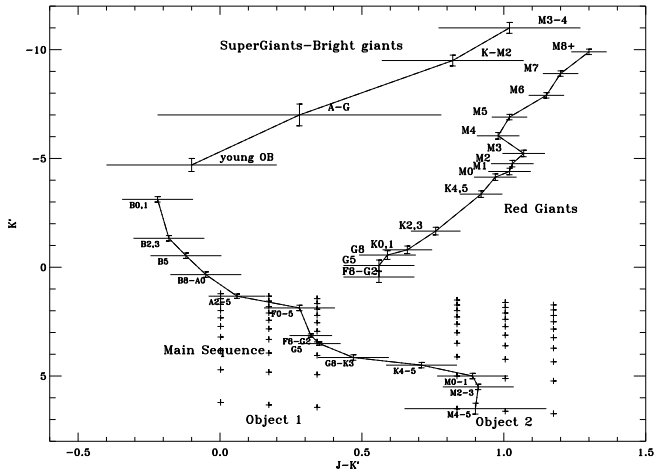,angle=90.,width=9.cm}}
\caption[]{\label{xtej1908_cmd} $J-K'$,  $M_{K'}$,absolute CMD diagram
computed for template main sequence stars \citep{ruelas-mayorga:1991}.
The range of locations for the $\xtejdnzh$ companion star in the
diagram has been derived from the absolute magnitudes of the two
possible counterparts, computed taking the minimum apparent magnitudes
on 2002 August 18th, and assuming that all the flux is produced from
the companion star. Three different values of the interstellar
extinction have been applied i.e. $A_{V}= 11.19, 12.9, 13.9$
magnitudes from right to left, respectively. $A_{V}=12.9$ magnitudes
corresponds to a column density $N_H = 2.3 \times 10^{22}$
\citep{woods:2002}, using the relation from \citet{predehl:1995}.  From
bottom to top the crosses correspond to a source distance increasing
from 1 to $10 \kpc$.  
According  to  the diagram,  in  the  case  of object  1,  the
companion  star would be  be an  intermediate/late type  main sequence
star of spectral  type between A and K, located between  3 and 10 kpc.
In the case of object 2,  the companion star would be a late-type main
sequence star of  spectral type later than K, located  between 1 and 3
kpc.  }
\end{figure}

We show  in Figure \ref{xtej1908_cmd} a  theoretical $J-K'$, $M_{K'}$,
colour-magnitude    diagram   (CMD)    built   for    template   stars
\citep{ruelas-mayorga:1991}.  We  overplot the locations corresponding
to the two $\xtejdnzh$ candidate counterparts, with the corresponding
absolute magnitudes being computed for ten different values of
the distance, ranging from $1$ to $10 \kpc$, and dereddened for three
different values of the hydrogen column density around the value of 
$N_H = 2.3 \times 10^{22} \cmmoinsdeux$ measured by \citet{woods:2002}.


Since each derived value of the absolute magnitude represents an upper
limit, we can not accurately determine the distance of the source from
the locations  of their candidate  counterparts on the  CMD.  However,
the  spectral type  can be  determined by  the $J-K'$  colour,  which is
independent  of  the  source  distance. Therefore,  according  to  the
diagram, the companion star of  object 1 would be an intermediate/late
type main  sequence star, probably of  spectral type between  A and K,
and at a distance between 3 and  10 kpc.  In the case of object 2, the
companion star  would be a  late-type main sequence star,  probably of
spectral type later than K, and at a distance between 1 and 3 kpc.

The derived  constraints  on the  spectral
classification  of the companion  star are  fully consistent  with the
non-detection at $R \ge 23$ \citep{wagner:2002} and at $ I \ge
22$ \citep{garnavich:2002}, since  a K or M type star at a distance $\ge
1 kpc$ would yield, for the expected absorption, $V \ge 30$. 
Near-infrared observations are therefore of prime importance
to reveal highly absorbed sources.  

However, taking into account the peak bolometric flux of $\xtejdnzh$,
\citet{in'tzand:2002b} have yielded a lower limit to the distance of 3 kpc
(and even 6 kpc if the compact object is a black hole, with a minimum
mass of 3 $M_\odot$). This makes it unlikely that object 2 is the
actual counterpart of $\xtejdnzh$.  Therefore, our analysis of the CMD
favours, as does the astrometry, object 1 as the most likely
counterpart.


\section{Conclusion}

We have reported on CFHT adaptive optics observations of the black
hole candidate $\xtejdnzh$ which showed that the previously reported
counterpart \citep{chaty:2002c} was in fact a blend of two objects. A
new and more accurate astrometric analysis provides qualitative
evidence that the brighter of the two (called object 1) is a more
likely counterpart. Accurate photometry performed on four sets of
observations spanning 4 months have shown a decrease of the NIR flux
for one (or both) of the two candidates, consistent with the X-ray
decay, showing that one of the two is almost certainly the real
counterpart of $\xtejdnzh$.
After dereddening and assuming that the flux represents upper limits
to the contribution of the companion star, we derived that the
companion star could be i) either an intermediate/late type main
sequence star of spectral type between A and K, located between 3 and
10 kpc, or ii) a late-type main sequence star of spectral type later
than K, located between 1 and 3 kpc. However, since a lower limit to
the distance of 3 kpc has been derived from the peak bolometric flux
of $\xtejdnzh$ by \citet{in'tzand:2002b}, both colors and magnitudes,
as well as the astrometry, would rule out object 2 and favour object 1
as the most likely counterpart of the X-ray source. 

Further near-infrared observations are needed, especially during
the next outburst of the source, in order to monitor the variability of both
candidate counterparts and eventually confirm (or not) our identification
of object 1 as the counterpart of $\xtejdnzh$. Getting some 
spectroscopy, although difficult because of faint and blended candidates, 
should be feasible with next generation instruments, and would finally 
better constrain both the spectral type and distance of the companion star.

\section{acknowledgements}
This work is partially supported through Agenzia Spaziale Italiana
(ASI), Ministero  dell'Istruzione, Universit\`a e Ricerca Scientifica e
Tecnologica (MIUR -- COFIN), and Istituto Nazionale di Astrofisica
(INAF) grants.
%
This research has made use of NASA's Astrophysics Data System.
Bibliographic Services and quick-look results provided by the ASM/RXTE team.


\end{document}